\documentclass[runningheads]{llncs}
\usepackage{multirow}
\usepackage{graphicx}
\usepackage{ulem}
\usepackage{xcolor}
\usepackage{array}
\usepackage{booktabs}
\usepackage{caption}
\usepackage{subcaption}
\usepackage{rotating}
\usepackage{makecell}
\usepackage{url}
\usepackage{fmtcount}

\begin{document}
\title{A Benchmark of Nested Named Entity Recognition Approaches\\in Historical Structured Documents%
%\thanks{Supported by ANR SoDUCo.}
}
\titlerunning{A Benchmark of Nested NER Approaches in Historical Documents}
\author{
%(Anonymous submission)
Solenn Tual\inst{1}\orcidID{0000-0001-8549-7949} \and
N. Abadie\inst{1}\orcidID{0000-0001-8741-2398} \and
J. Chazalon\inst{2}\orcidID{0000-0002-3757-074X} \and
B. Duménieu\inst{3}\orcidID{0000-0002-2517-2058} \and
E. Carlinet\inst{2}\orcidID{0000-0001-5737-5266}
}
%
%\authorrunning{S. Tual et al.}
\authorrunning{\mbox{}}
\institute{%
\mbox{}
 1
LASTIG, Univ. Gustave Eiffel, IGN-ENSG, F-94160 Saint-Mandé, France\\
\email{\{solenn.tual,nathalie-f.abadie\}@ign.fr}\\
 \url{https://www.umr-lastig.fr}
\and
 2
EPITA Research Laboratory (LRE), Le Kremlin-Bicêtre, France\\
\email{\{edwin.carlinet,joseph.chazalon\}@lrde.epita.fr}\\
\url{https://lrde.epita.fr}
\and
 3
CRH-EHESS, Paris, France\\
\email{bertrand.dumenieu@ehess.fr}\\
 \url{http://crh.ehess.fr/index.php?5206}\\
}
\maketitle % typeset the header of the contribution
\begin{abstract}
%The abstract should briefly summarize the contents of the paper in
%15--250 words.
%1200 chars max
Named Entity Recognition (NER) is a key step in the creation of structured data from digitised historical documents. 
Traditional NER approaches deal with flat named entities, whereas entities often are nested. For example, a postal address might contain a street name and a number. This work compares three nested NER approaches, including two state-of-the-art approaches using Transformer-based architectures. We introduce a new Transformer-based approach based on joint labelling and semantic weighting of errors, evaluated on a collection of 19\textsuperscript{th}-century Paris trade directories. We evaluate approaches regarding the impact of supervised fine-tuning, unsupervised pre-training with noisy texts, and variation of IOB tagging formats.
Our results show that while nested NER approaches enable extracting structured data directly, they do not benefit from the extra knowledge provided during training and reach a performance similar to the base approach on flat entities. Even though all 3 approaches perform well in terms of F1 scores, joint labelling is most suitable for hierarchically structured data. Finally, our experiments reveal the superiority of the IO tagging format on such data.

\end{abstract}
%
% TODO figure générale
%
\begin{figure}[tb]
\centering
\includegraphics[width=0.7\textwidth]{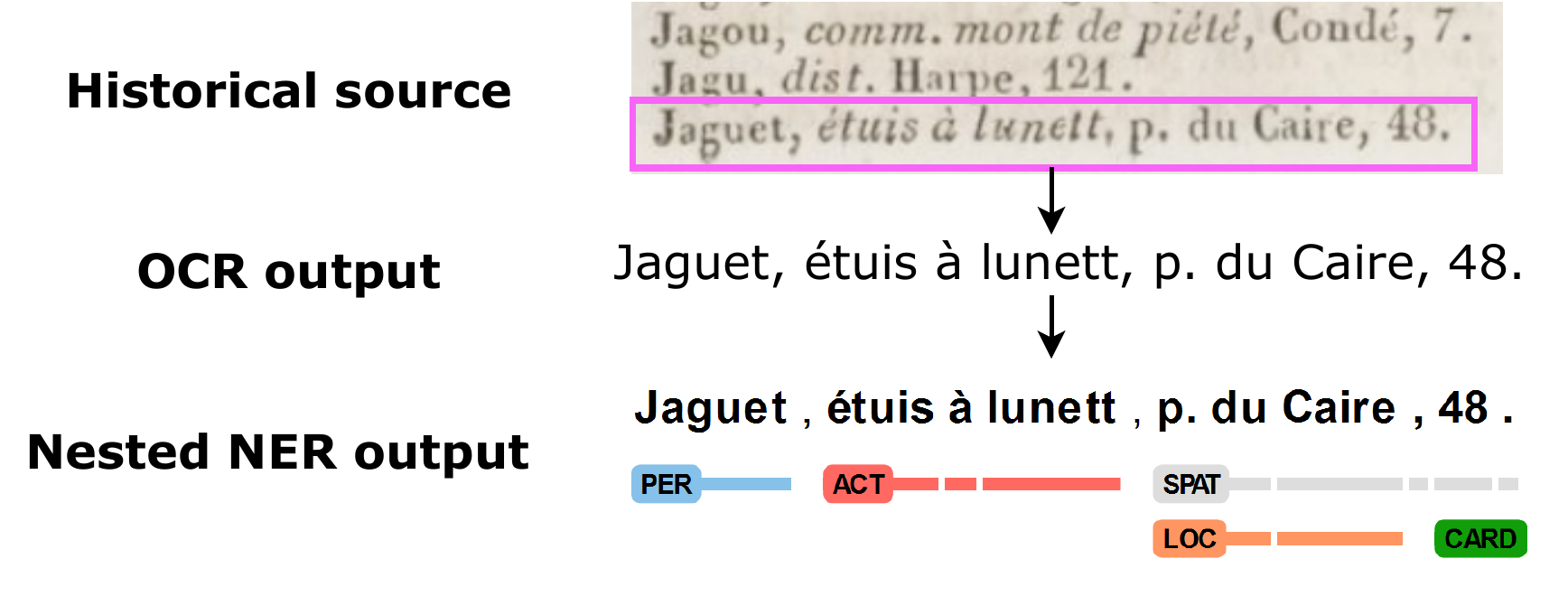}
\caption{%
% Overview of the digitised document, OCR result, and structured output associated with the Nested Named Entity Recognition task on historical documents
Overview of the Nested Named Entity Recognition task on historical documents, where multiple entities can be assigned to the same text span.
}\label{fig1}
\end{figure}

\section{Introduction}

Named entity recognition (NER) is a classic natural language processing task used to extract information from various types of textual documents. A named entity is an entity with a specific meaning, such as a person, an organisation, or a place. \cite{finkel_nested_2009} point out that entities are often nested. State-of-the-art approaches are divided into flat NER approaches, which associate a span with zero or one entity type, and nested named entity recognition approaches, which can associate a given span with zero, one, two or more entity types. A nested entity is thus composed of one or more entities. Base entities can be associated to create more complex entities. Nested named entities are also referred to as tree-structured entities \cite{dinarelli_models_2011}, structured entities \cite{wajsburt_extraction_2021} or hierarchical named entities \cite{marinho_hierarchical_2019}.

We focus on the particular case of nested named entities in historical documents with an application to the 19\textsuperscript{th} century trade directories of Paris, France. Each directory contains tens of thousands of entries organised into lists in alphabetical order. Each entry is a highly structured non-verbal short text (see Figure \ref{fig1}). It refers to a company or a person name, followed by a description of varying complexity that often include details on his/her professional activities, and ends with one or several addresses. Additional information like military or professional awards are sometimes given. The directories were published by multiple editors between the late 18\textsuperscript{th} century and the second half of the 19\textsuperscript{th} century. 

Similar work aiming to extract information from historical directories exists in the literature. They use flat NER approaches complete by a post-processing stage to create hierarchical entities. \cite{bell_automated_2020} used directories published between 1936 and 1990 to build a database of gas stations in the city of Providence, Rhode Island, United States. The pipeline consists of the following steps: an image processing block, an Optical Character Recognition (OCR) block, and a named entity recognition block performed using a rule-based approach. To limit the impact of OCR errors, the authors clean the texts before the NER stage. Structured entities, such as addresses, are built up in post-processing. \cite{albers_perks_2022} developed a similar pipeline that was used to extract data from the Berlin directories of 1880. \cite{abadie_das_2022} introduce a deep learning approach in their pipeline to perform NER on noisy entries extracted from the Paris trade directories. Structured postal addresses are also built in post-processing before geocoding.

We focus on nested NER in historical sources to extract structured entities without rule-based post-processing. It requires taking into account the main characteristics of the books. First, entries contain two levels of entities. Top-level entities are built using bottom-level entities. For example, an address consists of two or three other types of entities: a street name, a number and sometimes a geographical feature type (e.g. \textit{depôt} - warehouse). All of these spans are not formally linked as an address with flat-NER approaches, while they are linked with nested NER approaches. These approaches provide a more complex semantic view of directories, which improves searchability. Second, directories' entries do not have the same structure throughout the corpus: the entities' enumeration pattern varies. Finally, directory entries are extracted from digitised documents using OCR models. This requires dealing with noisy texts at the NER stage. The amount of noise changes depending on the quality of the original print, the conservation status of historical sources, and the resolution of the digitised images. OCR errors are erroneous, missing, or extra characters. These multiple patterns of entities and the noisy entries encountered in the collection of directories make the use of rule-based methods inappropriate.

In this article, we focus on the evaluation of several nested named entity recognition approaches. The main contributions are new nested named entity datasets using Paris trade directories built on an existing flat NER dataset, and a benchmark of three nested named entity recognition approaches using transfer-learning models fine-tuned on a ground-truth dataset and on a noisy dataset. We use two pre-trained BERT-based models: a state-of-the-art French model and this same model pre-trained on domain-specific noisy examples to assess the impact of unsupervised training on the performances. We also compare two tagging formats to evaluate their influence on NER performance in highly structured texts.

 This paper is organised as follows: (i) a state-of-the-art of nested NER approaches, OCR noise, and tagging strategies on NER task; (ii) a presentation of the evaluated approaches ; (iii) an introduction of our datasets, experiments and metrics ; (iv) an evaluation of our approaches from both quantitative and qualitative point of view regarding the effects of the tagging strategy and unsupervised pre-training on transfer-learning-based models.

\section{Related works}

We want to produce searchable structured data using historical sources, such as the Paris trade directories. This paper aims to deal with properties of this corpus' writings: short and highly structured texts, including OCR noise, which contained nested-named entities.

\subsection{Nested Named Entity Recognition approaches}

The survey presented in \cite{wang_nested_2022} identifies groups of nested NER approaches, among them the rule-based approaches, the layered-based approaches, and the region-based approaches. While the first one mostly relieves early methods, the others often fall under supervised learning, which includes traditional machine learning and deep learning approaches. 
%Il y a d'autres types d'approches comme :
% Hypergraph-based
% Transition-based

The hand-crafted rule-based approaches aim to recognise patterns in documents. These methods require a high level of data and linguistics expertise to identify and design extraction rules, as explained by \cite{shen_effective_2003}. Most of the time, they involve searching for vocabulary in gazetteers or dictionaries as in flat-NER approaches \cite{nadeau_survey_2007}. Layered approaches handle the nested NER task using cascades of flat NER layers. Each layer is trained to recognise entities of a given group according to their depth level in the dataset. \cite{jia_nested_2021} suggests stacking predictions made by independent layers for each entity level. Models often use n-level outputs to improve the recognition of n+1 level entities such as \cite{ju_neural_2018} with a Layered BiLSTM+CRF model or \cite{wajsburt_effect_2021} with a BERT+CRF model. Region-based approaches use two-step models to detect candidates and classify them. There are multiple entity listing strategies: enumeration strategies (such as n-gram) or boundaries-based strategies that aim to identify the first and last tokens of potential entities \cite{zheng_boundary-aware_2019}.
In addition to these approaches, \cite{wang_nested_2022} introduce various methods which have in common the use of a unique tag created by concatenating multiple labels from the nested entities. \cite{agrawal_bert-based_2022} propose an intuitive approach using this unique multilevel label to fine-tune a flat NER BERT-based model. Deep learning methods outperform state-of-the-art approaches most of the time according to \cite{wang_nested_2022,li_survey_2022}.

\subsection{Named Entity Recognition on noisy inputs}

With the rise of the digitisation of historical documents, the need to extract and structure the information they contain has increased dramatically. Named Entity Recognition is a useful way to produce a high-level semantic view of these documents. Performing this task on historical documents has its own peculiarities that need to be taken into account. \cite{ehrmann_named_2021} lists specific challenges associated with the NER task in historical documents. Most of them are processed with automatic tools such as OCR or Handwritten Text Recognition (HTR), which are likely to misread some characters. Thus, it involves dealing with noisy inputs in the NER stage. \cite{dinarelli_tree-structured_2012} suggests pre-processing texts before a NER stage. They process a dataset of French newspapers of the 19\textsuperscript{th} century and underline the impact of OCR errors on the recognition of tree-structured named entities.
\cite{abadie_das_2022} study the effects of OCR noise on flat NER with transfer learning models in 19\textsuperscript{th} century directory entries. According to \cite{li_survey_2022}, they made an unsupervised pre-training of the French BERT-based model CamemBERT with domain-specific texts provided by OCR and fine-tuned this model with noisy annotated examples. It improved the tolerance of the NER model to noise, and its performances have been improved compared to a non-specialised model.

\subsection{Labels for NER on highly structured documents}

In the named entity recognition task, each word (or token) of an entity is associated with a label, also called a tag. Labels follow specific writing formats. The most widely used format is the Inside-Outside-Beginning (IOB) model introduced by \cite{ramshaw_text_1995}. In practise, these formats define prefixes in front of each class name to specify the position of the token in the entity. The table \ref{tab:labelsiob} presents tags associated with a directory entry in the IO and IOB2 formats. These tags are required for the training stage. There are several variants of this tagging standard: IO, IOB2, IOE, IOBES, BI, IE, or BIES. The choice of tag format has an impact on the performance of named entity recognition models. In their survey, \cite{alshammari_impact_2021} shows that IO gives the best results on fully developed and non-noisy texts among a set of seven annotation formats derived from the IOB format.

\begin{table}[tb]
    \centering
    \caption{IO and IOB2 labels example for \textit{Aubery je. r. Quincamp. pass. Beaufort.} entry (Cambon almgene, 1841). Entity labels are defined in table \ref{tab:entities}.}
    \label{tab:labelsiob}
    \begin{tabular}{|p{2cm}|p{1.5cm}|p{1,5cm}|}
       \hline
       \textbf{Token} & \textbf{IO} & \textbf{IOB2}  \\
       \hline
       Aubery & I-PER & \textbf{B}-PER \\
       \hline
       je & I-PER & I-PER \\
       \hline
       . & I-PER & I-PER \\
       \hline
       r & I-LOC & \textbf{B}-LOC \\
       \hline
       . & I-LOC & I-LOC \\
       \hline
       Quincamp & I-LOC & I-LOC \\
       \hline
       . & I-LOC & I-LOC \\
       \hline
       pass & I-LOC & \textbf{B}-LOC \\
       \hline
       . & I-LOC & I-LOC \\
       \hline
       Beaufort & I-LOC & I-LOC \\
       \hline
       . & O & O \\
       \hline
    \end{tabular}
\end{table}

\subsection{Conclusion}

The following benchmark concentrates on the three main properties of our dataset: nested entities, noisy inputs, and highly structured texts. We compare three nested named entity recognition approaches based on deep learning models to recognise complex entities in directories entries. We measure the impact of two tagging formats on entity recognition in short texts with repetitive patterns of entities. We evaluate the effects of domain adaptation of BERT-based models on noisy inputs. Finally, we aim to evaluate the contribution of training with nested entities on the flat named entity recognition.

\section{Considered Nested NER approaches}

In this section, we present the three evaluated approaches. We focus on intuitive, low-complexity and high-performance approaches using transfer learning models. We implement two state-of-the-art proposals and introduce a new approach that incorporates hierarchical information into training.

\subsection{State-of-the-art nested NER approaches}

%We considers a region-based approach implemented in SpaCy NLP library proposes a two-step pipeline with a suggester, which identifies span boundaries, and a classifier to identify span types. We use the SpanFinder component as a suggestion. This tool outperforms traditional N-Gram enumeration approaches. It suggests the boundaries of the range by tagging the potential start and stop tokens of elements that may overlap. Spans are then classified with a CNN model. This pipeline does not require GPU for small datasets. \cite{spanfinder} \nathalie{Retrouver le dépôt où j'ai récupéré le code Spacy et le mettre en notre de bas de page?}

\subsubsection{Independent NER Layers (abbreviated as [M1]).} \cite{jia_nested_2021}'s approach proposes to fine-tune a pre-trained BERT-based model for each level of entities in the dataset. Tags examples are given in table \ref{tab:tags}. Each model is called a layer and is completely independent of the others. The predictions made by each layer are merged to obtain nested entities. This approach reduces the risk of error propagation: a level 2 entity can be detected even if level 1 is wrong. However, the independence of the NER layers does not allow us to control the combination of the layers' predictions. The assignment of a class to a span at level N does not favour or disfavour the assignment of another class at level N+1.

\subsubsection{Transfer learning approach using joint labelling [M2].} \cite{agrawal_bert-based_2022} propose a BERT-based transfer learning approach using joint labelling. For each token, the authors create a complex label composed of the labels of each depth level of the nested dataset, as described in table \ref{tab:tags}. This label is called \textit{joint label}. A single pre-trained BERT-based model is fine-tuned with these annotated data. The nested levels are processed simultaneously as a single level. It implies an increase in the number of labels used for training.

\begin{table}[tb]
    \centering
    \caption{IOB2 tags used to train M1 and M2. The entry used as an example is \textit{Dufour (Gabriel), bookseller, Vaugirard Street, 7.} Entity labels are defined in table \ref{tab:entities}.}
    \begin{tabular}{|p{1.8cm}|p{2.5cm}p{2.5cm}|p{3.5cm}|}
       \hline
       \textbf{Token} & \textbf{Level-1 [M1]} & \textbf{Level-2 [M1]} & \textbf{Joint label [M2]} \\
       \hline
       Dufour & B-PER & O & B-PER+O\\
       \hline
       ( & I-PER & O & I-PER+O\\
       \hline
       Gabriel & I-PER & O & I-PER+O\\
       \hline
       ) & I-PER & O & I-PER+O\\
       \hline
       , & O & O & O+O\\
       \hline
       libraire & B-ACT & O & B-ACT+O\\
       \hline
       , & O & O & O+O\\
       \hline
       r & B-SPAT & B-LOC & B-SPAT+B-LOC\\
       \hline
       . & I-SPAT & I-LOC & I-SPAT+I-LOC\\
       \hline
       de & I-SPAT & I-LOC & I-SPAT+I-LOC\\
       \hline
       Vaugirard & I-SPAT & I-LOC & I-SPAT+I-LOC\\
       \hline
       , & I-SPAT & O & I-SPAT+O\\
       \hline
       7 & I-SPAT & B-CARDINAL & I-SPAT+B-CARDINAL\\
       \hline
    \end{tabular}
    \label{tab:tags}
\end{table}

\subsection{A hierarchical BERT-based transfer-learning approach using joint labelling [M3]}

\begin{figure}[tb]
     \centering
      \includegraphics[width=0.9\textwidth]{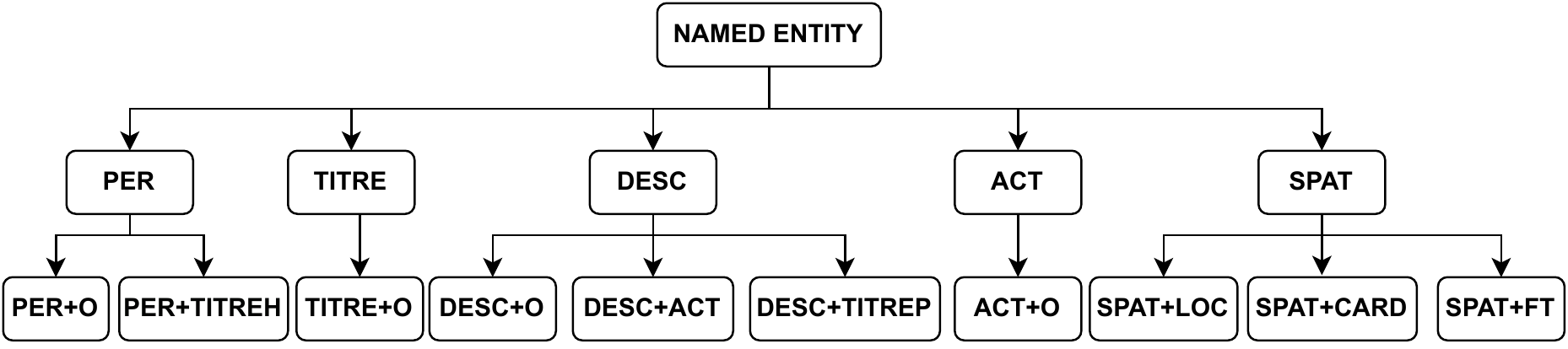}
      \caption{Multi-level tree representation of IO tags used by the Hierarchical Loss Function called in M3 to compute the semantic distance between labels.}
      \label{fig:m3tree}
\end{figure}

We propose a new version of the BERT-based joint labelling approach of \cite{agrawal_bert-based_2022}, inspired by computer vision work on segmentation and classification. \cite{bertinetto_making_2020} note that the loss function used for classification tasks considers all errors with the same weight, regardless of the semantic proximity between classes. Therefore, they propose to take into account the semantic distance between classes in the training and evaluation process of the models by modifying the loss function. 
We replace the Categorical Cross Entropy Loss implemented in the original CamemBERT model with the Hierarchical Cross Entropy Loss developed by \cite{bertinetto_making_2020}. This function incorporates the semantic distance calculated using a tree that defines the class hierarchy. Each joint label corresponds to a leaf of the tree, as shown in Figure \ref{fig:m3tree}. The errors are weighted according to the distance between the expected class and the predicted class in the tree. For example, if the model predicts a \textit{SPAT+FT} entity instead of a \textit{SPAT+LOC} entity, it is a \textit{better error} than if it predicts a \textit{DESC+O} entity because the labels are semantically closer in the first case than in the second according to the tree. Using IOB2 tags, a tree level is added to distinguish the positional prefixes that make up the labels of each class. The loss function of \cite{bertinetto_making_2020} allows the weighting of the branches of the tree to increase the impact of some classification errors. This last possibility has not been treated in our work. 
%\nathalie{Ajouter des exemples? Expliquer que dans le cas de bertinetto, le but est de calculer la distance sémantique sur une hierarchie de type isA alors que nous avons plutôt une hierarchie partOf?}

\section{Evaluation}

In this section, we describe our datasets, our experiments and their associated parameters including the tagging formats, the pre-trained models and the hyperparameters used to fine-tune them. Finally, we define our metrics.

\subsection{Datasets}

We have produced two annotated datasets using Parisian trade directories: the ground-truth dataset with non-noisy entries and a real-world dataset with the corresponding noisy OCR outputs. The use of these two datasets allows us to compare the performance of the approaches on both clean and noisy texts.

\subsubsection{Source documents: Paris trade directories from 1798 to 1854.}

\begin{figure}[tb]
     \centering
     \begin{subfigure}[b]{0.35\textwidth}
         \centering
         \includegraphics[width=\textwidth]{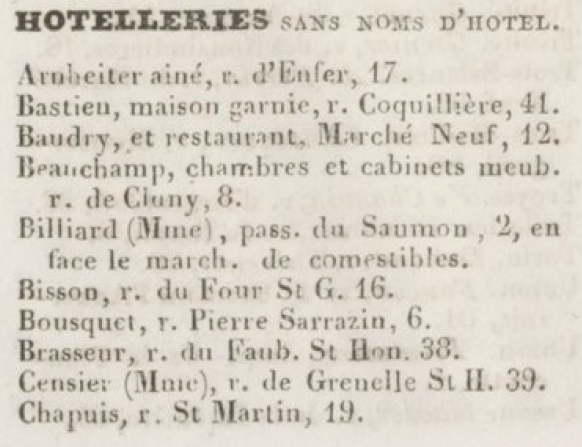}
         \caption{List ranked by activity}
         \label{fig:listbyjob}
     \end{subfigure}
     \hfill
     \begin{subfigure}[b]{0.60\textwidth}
         \centering
         \includegraphics[width=\textwidth]{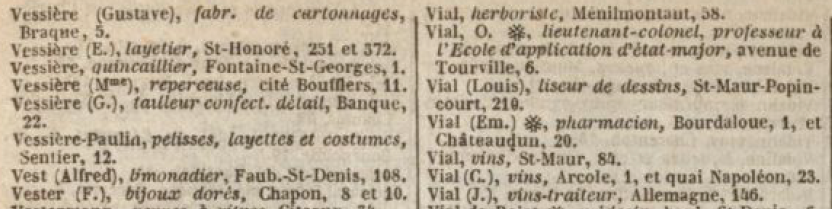}
         \caption{List ranked by name}
         \label{fig:listbyname}
     \end{subfigure}
        \caption{Extracts from Cambon Almgène (\ref{fig:listbyjob}, 1839) and Didot-Bottin (\ref{fig:listbyname}, 1874) directories.}
        \label{fig:directoriesextracts}
\end{figure}

Our aim is to detect nested entities in entries published during the 19\textsuperscript{th} century. They were printed over a long period of time by several editors using different printing techniques. The layout, content, length, and typography (font, case, abbreviations, etc.) of the entries vary greatly, as illustrated in figure \ref{fig:directoriesextracts}. Some directories organise the information into several lists, sorted by name, activity, or street name. This means that there are many different types of entities to find. Some types of entities are common, while others are sparcer. Several independent organisations have digitised directories of varying quality. This results in more or less clean OCR results.

\subsubsection{Directories datasets}
\begin{figure}[tb]
\centering
\includegraphics[width=0.55\textwidth]{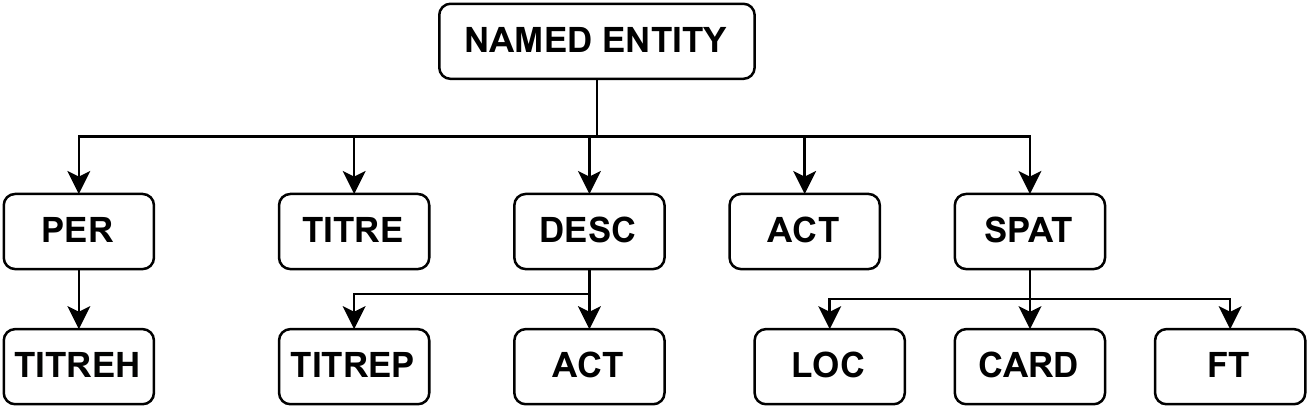}
\caption{\textit{Part-of} hierarchy describing the nested entities.} \label{fig_hierarchy}
\end{figure}

We produced two new NER datasets  from the public dataset of~\cite{abadie_dataset_2022}. Both datasets are created from the same 8765 entries from 78 pages 
 chosen in 18 different directories published from 1798 to 1861. The text is extracted using Pero-OCR. 
 The first dataset is created by first correcting the text predicted by Pero-OCR and then annotating the ground-truth named entities.  
To create the second dataset, also known as the \textit{real world} dataset, we first align the text produced by Pero-OCR with the ground-truth text by means of tools implemented as part of Stephen V. Rice's thesis~\cite{neudecker2021,santos2019}. Lastly, the named entities spans and their associated labels are projected from the first dataset text to the Pero-OCR text. Entries for which no named entities could be projected are also removed from the real world dataset, which in the end has 8445 entries.
For our experiments, we reduce the ground-truth dataset to the 8445 entries that also are in the noisy dataset. The ground-truth dataset is used to perform the evaluation of the approaches without the impact of the OCR and compare its result with one of the models trained with noisy data.
Entries have been randomly selected in several directories to represent the wide variety of typographic and pattern types to be learned. There are 10 types of entities described in table~\ref{tab:entities}. The maximum level of entities is 2. The entity hierarchy is described in figure~\ref{fig_hierarchy}. It's a \textit{Part-Of} hierarchy: combined level-2 entities form a level-1 entity.

\begin{table}[tb]
\centering
\caption{Named entity types, description, and count in the ground-truth directories dataset (8445 entries).}
\label{tab:entities}
    \begin{tabular}{|l|l|l|l|}
        \hline
       \textbf{Entity} & \textbf{Level} & \textbf{Description} & \textbf{Count}\\
       \hline
       PER & 1 & Person(s) or business name. & 8441 \\
       ACT & 1 or 2 & Person or company's activities & 6176 \\
       DESC & 1 & Complete description. & 371 \\
       SPAT & 1 & Address & 8651 \\
       TITREH & 2 & Military or civil title relative to company's owner & 301 \\
       TITREP & 2 & Professional rewards & 94 \\
       TITRE & 1 & Other title. & 13 \\
       LOC & 2 & Street name & 9417 \\
       CARDINAL & 2 & Street number & 8416 \\
       FT & 2 & Kind of geographic feature & 76 \\
       \hline
    \end{tabular}
\end{table}

\subsection{Experiments summary}

\subsubsection{Nested NER on noisy texts}

The first axis of our work focusses on the comparison of nested NER approaches to directory entries. We evaluate the ability of these approaches to detect all entities regardless of hierarchy and, in contrast, to associate the correct type of level 1 entity with any level 2 entity. We include in these experiments the use of two pre-trained transformer models and two tagging formats. The robustness of the approaches to OCR noise is given particular attention.

\subsubsection{Flat NER vs. Nested NER}

\begin{figure}[tb]
\centering
\includegraphics[width=0.9\textwidth]{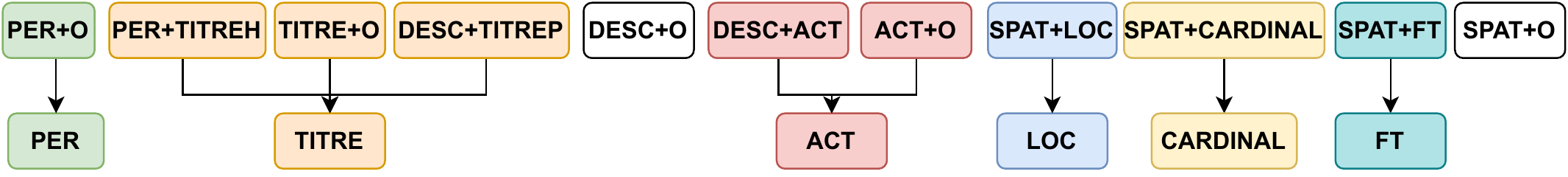}
\caption{Mapping between nested NER entity types and flat NER entity types.} \label{fig:mappingdas}
\end{figure}

The second axis of this paper is to assess the impact, positive or negative, of the use of complex labels on the recognition of flat named entities. We created flat NER datasets (clean and noisy) using our tree-structured annotations. We map the joint labels to the flat labels described by \cite{abadie_das_2022} as described in figure \ref{fig:mappingdas} and reimplement their flat NER experiment. We do not use \cite{abadie_dataset_2022} due to the changes we made during our tree-structured annotation phase. We also map our predicted nested entities to the corresponding flat entity types during the nested NER experiments to compare F1-score values.

\subsection{Tagging formats}

The tagging formats retained for this benchmark are IO and IOB2. The IO format is the least restrictive IOB-like tagging model. If a token is part of an entity, it is tagged \textit{I-class}, otherwise it is tagged \textit{O}. However, it does not allow the model to distinguish two entities of the same type that follow each other without separation (see the example in Table \ref{tab:labelsiob}). This pattern sometimes appears in our historical records. We also use the IOB2 format. This choice is based on the hypothesis that tagging each first token of an entity can help to divide a sequence of entities of the same type (as two successive \textit{Addresses}). This involves creating two tags for each entity type: \textit{B-class} and \textit{I-class}.

\subsection{Pre-trained BERT-based models}

We fine-tuned two BERT-based pre-trained models to evaluate the impact of unsupervised domain pre-training on performances. We choose CamemBERT \cite{martin_camembert_2020}, a French BERT-based pre-trained model that has been fine-tuned for NER (CamemBERT NER) and a domain-specific version of this model pre-trained with OCR outputs from Paris trade directories \cite{abadie_das_2022}.
All the experiments are led with hyperparameters presented in table \ref{tab:hyperparameters}.

\begin{table}[tb]
\centering
\caption{Hyperparameters used for fine-tuning}
\label{tab:hyperparameters}
\begin{tabular}{|l|l|lll}
\cline{1-2} \cline{4-5}
\textbf{Learning rate} & 1e-4                   & \multicolumn{1}{l|}{} & \multicolumn{1}{l|}{\textbf{Callback patience}}     & \multicolumn{1}{l|}{5}        \\ \cline{1-2} \cline{4-5} 
\textbf{Weight decay}  & 1e-5                   & \multicolumn{1}{l|}{} & \multicolumn{1}{l|}{\textbf{Evaluation strategy}}   & \multicolumn{1}{l|}{Steps}    \\ \cline{1-2} \cline{4-5} 
\textbf{Batch size}    & 16                     & \multicolumn{1}{l|}{} & \multicolumn{1}{l|}{\textbf{Max steps number}}      & \multicolumn{1}{l|}{5000}     \\ \cline{1-2} \cline{4-5} 
\textbf{Optimizer}     & AdamW                  & \multicolumn{1}{l|}{} & \multicolumn{1}{l|}{\textbf{Metric for best model}} & \multicolumn{1}{l|}{F1-Score} \\ \cline{1-2} \cline{4-5} 
\textbf{Seed}          & Run number &                       & \textbf{}                                           &                               \\ \cline{1-2}
\end{tabular}%
\end{table}

\subsection{Metrics}

We use the \textit{seqeval} \cite{seqeval} library to evaluate the performance of each approach. It is well suited to the evaluation of natural language processing tasks, including sequence labelling. The tool supports the IO and IOB2 tag formats. The metrics used for the evaluation are precision, recall, and the F1-score.
To compute these values, the tool first gathers tokens of the same class that follow each other. These groups of tokens constitute the predicted entities that are aligned with the ground-truth. Any difference in the boundary of the entity or its class, at any level, will be considered as an error. 

\begin{figure}[tb]
\centering
    \includegraphics[width=0.9\textwidth]{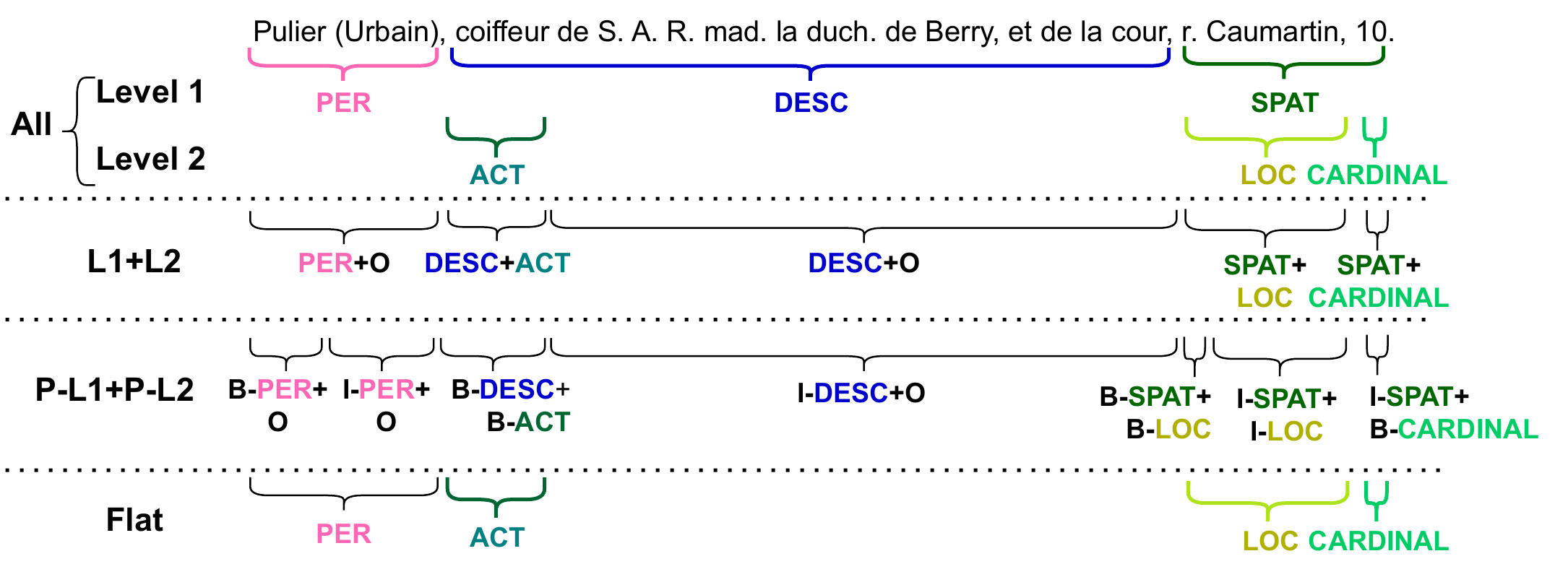}
    \caption{Entities used for each evaluation metric.} 
    \label{fig:evaluationboundaries}
\end{figure}

We look at the performance of approaches in nested NER. We calculate the F1-score for the spans illustrated in Figure \ref{fig:evaluationboundaries}. We can split them in two groups. First, we evaluate the ability of approaches to recognise entities independently of their structure. \textbf{All} is the global measure on entities of both levels, \textbf{Level 1 and Level 2} are the performance measure on entities of each entity level. Second, we evaluate the ability of models to deal with entities' hierarchy. \textbf{L1+L2} evaluate the fact that a Level-2 entity is part of a well-predicted Level-1 entity. \textbf{P-L1+P-L2} includes the evaluation of positional prefixes to the previous definition of \textit{L1+L2}. For IO tagging format, \textit{P-L1+P-L2} values are equal to \textit{L1+L2} score and for IOB2 tagging format, \textit{P-L1+P-L2} score compare expected and predicted entity types and IOB-like prefixes. Finally, the value \textbf{Flat} is the F1-score computed on the flat NER types mapped from \textit{L1+L2} predictions (see figure \ref{fig:mappingdas}).

\section{Results}

In this last section, we present the results of our experiments from a quantitative and a qualitative point of view. 

\subsection{Performances}

We run the three nested NER approaches and a state-of-the-art flat NER approach with two main goals : (i) compare nested-NER approaches regarding the impact of tagging formats and unsupervised domain-specific pre-training of BERT-based models and (ii) evaluate the performance of nested-NER approaches regarding flat-NER ones. F1-Score values are measured on the 1685 entries of the test subset and are the mean of 5 runs with fixed seeds.

\subsubsection{Nested-NER experiments results}
Table \ref{tab:globalmetrics} presents the F1-score measured for real-world (noisy OCR) and ground-truth datasets. Most of the time, the IO tagging format gives the best results on both datasets. Unsupervised pre-training of the BERT-based models with domain examples increases the results in most cases and benefits the noisy dataset. The three approaches produce close results. On the \textit{All} entities F1-score value, metrics are included between 95.6\% and 96.6\% in the ground-truth dataset and between 93.7\% and 94.3\% in the noisy dataset. The independent NER layers approaches (M1) using IO tags outperforms all tests on both datasets on \textit{All} and \textit{L1} entities, while the joint labelling approach with hierarchical loss give the best scores on \textit{L1+L2} and \textit{P-L1+P-L2} spans. \textit{L2} best-score is reached with M1 on the ground-truth dataset and M3 on the noisy dataset. 

M1 is trained to recognise entities of each independent layer. Our results show that it optimizes scores of non-hierarchical features whereas M3, which is trained to maximise \textit{P-L1+P-L2} spans recognition, outperforms all approaches on hierarchical entity detection. We focus on \textit{L1+L2} and \textit{P-L1+P-L2} values which should be favoured to produce structured data. Thus, to recognise nested named entities in noisy inputs, our proposed M3 approach with domain-adapted BERT-based model fine-tuned with noisy annotated examples and tagged with IO labels, is the best model to use. Note that the total number of parameters trained during the fine-tuning of the first approach is twice as large as for approaches 2 and 3 for a lower number of tags to be learned.

\begin{table}[tb]
\centering
\caption{F1 score measured for each approach, dataset, pre-trained models and tag formats (mean of 5 runs).}
\label{tab:globalmetrics}
\begin{tabular}{lllcccccc}
\toprule
 &  & \textbf{Model and tags} &  \textbf{All} &  \textbf{L1+L2} &   \textbf{L1} &   \textbf{L2} &  \textbf{P-L1+P-L2} &  \textbf{Flat}       \\
\midrule
\multirow{12}{*}{Ground-truth} & \multirow{4}{*}
    {M1} & CmBERT IO & 96.5 &   95.7 & 96.0 & \textbf{97.0} &       95.7 &  97.0 \\
    &    & CmBERT IOB2 & 96.2 &   95.6 & 95.8 & 96.8 &       95.7 &  96.1 \\
    &    & CmBERT+ptrn IO & \textbf{96.6} &   95.9 & \textbf{96.3} & \textbf{97.0} &       95.9 &  \textbf{97.3} \\
    &    & CmBERT+ptrn IOB2 & 96.0 &   95.2 & 95.5 & 96.6 &       95.3 &  95.3 \\
\cline{2-9}& \multirow{4}{*}
   {M2} & CmBERT IO & 96.2 &   96.2 & 95.6 & 96.9 &       96.2 &  96.7 \\
    &    & CmBERT IOB2 & 96.0 &   96.0 & 95.3 & 96.8 &       96.0 &  96.7 \\
    &    & CmBERT+ptrn IO & 96.3 &   96.1 & 96.0 & 96.7 &       96.1 &  96.9 \\
    &    & CmBERT+ptrn IOB2 & 96.1 &   96.0 & 95.8 & 96.4 &       96.1 &  96.9 \\
\cline{2-9}
    & \multirow{4}{*}
    {M3} & CmBERT IO & 96.3 &   96.2 & 95.8 & 96.9 &       96.2 &  96.8 \\
    &    & CmBERT IOB2 & 96.1 &   96.1 & 95.6 & 96.7 &       96.1 &  96.8 \\
    &    & CmBERT+ptrn IO & 95.8 &   95.8 & 95.2 & 96.7 &       95.8 &  96.4 \\
    &    & CmBERT+ptrn IOB2 & 96.3 &   \textbf{96.3} & 96.0 & 96.7 &       \textbf{96.3} &  97.0 \\
\cline{1-9}
\cline{2-9}
\multirow{12}{*}{OCR} & \multirow{4}{*}
    {M1} & CmBERT IO & 93.8 &   93.4 & 93.1 & 94.6 &       93.4 &  94.2 \\
    &    & CmBERT IOB2 & 93.5 &   92.9 & 93.1 & 94.0 &       93.1 &  92.7 \\
    &    & CmBERT+ptrn IO & \textbf{94.3} &   93.8 & \textbf{94.1} & 94.5 &       93.8 &  94.4 \\
    &    & CmBERT+ptrn IOB2 & 94.1 &   93.5 & 93.7 & 94.5 &       93.7 &  94.5 \\
\cline{2-9}
    & \multirow{4}{*}
    {M2} & CmBERT IO & 93.8 &   94.1 & 93.3 & 94.4 &       94.1 &  94.5 \\
    &    & CmBERT IOB2 & 93.8 &   94.2 & 93.2 & 94.5 &       94.3 &  94.7 \\
    &    & CmBERT+ptrn IO & 93.9 &   94.1 & 93.4 & 94.4 &       94.1 &  94.6 \\
    &    & CmBERT+ptrn IOB2 & 93.7 &   94.1 & 93.1 & 94.5 &       94.2 &  94.8 \\
\cline{2-9}
    & \multirow{4}{*}
    {M3} & CmBERT IO & 94.1 &   \textbf{94.4} & 93.5 & \textbf{94.8} &       \textbf{94.4} &  94.8 \\
    &    & CmBERT IOB2 & 93.5 &   93.9 & 92.9 & 94.3 &       94.0 &  94.6 \\
    &    & CmBERT+ptrn IO & 94.1 &   \textbf{94.4} & 93.6 & \textbf{94.8} &       \textbf{94.4} &  \textbf{94.9} \\
    &    & CmBERT+ptrn IOB2 & 93.7 &   94.0 & 93.1 & 94.4 &       94.2 &  94.8 \\
\bottomrule
\end{tabular}
\end{table}
\begin{table}[tb]
\centering
\caption{F1 score measured for each approach, pre-trained model and tag format (mean of 5 runs) on the ground-truth dataset for each entity type.}
\label{tab:classesmetricsref}
\resizebox{\textwidth}{!}{\begin{tabular}{llrrrrrrrrr}
\toprule
 & Model \& tags &  PER &  ACT &  DESC &  TITREH &  TITREP &  SPAT &  LOC &  CARD &   FT  \\
\midrule
\multirow{4}{*}
{M1} & CmBERT IO & 91.8 & 94.8 &  49.6 &    11.7 &    97.4 &  97.5 & \textbf{97.9} &      98.1 & 36.2 \\
   & CmBERT IOB2 & 90.4 & 94.0 &  43.3 &    22.5 &    97.3 &  97.7 & 97.6 &      96.5 & 51.5 \\
   & CmBERT+ptrn IO & \textbf{92.6} & \textbf{95.7} &  \textbf{53.5} &    50.5 &    97.2 &  97.6 & 97.6 &      \textbf{98.4} & 53.4 \\
   & CmBERT+ptrn IOB2 & 89.8 & 93.6 &  44.4 &    41.5 &    97.2 &  97.7 & 97.4 &      97.2 & 50.4 \\
\cline{1-11}
\multirow{4}{*}
{M2} & CmBERT IO & 90.6 & 94.5 &  47.1 &    39.9 &    97.5 &  97.3 & 97.2 &      97.5 & 58.7 \\
   & CmBERT IOB2 & 90.3 & 93.8 &  36.8 &    39.8 &    97.3 &  97.4 & 97.4 &      97.9 & 56.0 \\
   & CmBERT+ptrn IO & 90.1 & 94.6 &  47.7 &    \textbf{58.1} &    \textbf{97.6} &  98.2 & 97.3 &      \textbf{98.4} & 57.3 \\
   & CmBERT+ptrn IOB2 & 90.1 & 94.4 &  42.9 &    38.6 &    97.1 &  \textbf{98.4} & 97.3 &      97.9 & 62.3 \\
\cline{1-11}
\multirow{4}{*}
{M3} & CmBERT IO & 91.1 & 94.6 &  49.4 &    43.5 &    97.4 &  97.7 & 97.2 &      97.2 & 50.9 \\
   & CmBERT IOB2 & 90.8 & 94.3 &  44.8 &    40.7 &    97.3 &  97.3 & 97.4 &      95.8 & 51.8 \\
   & CmBERT+ptrn IO & 90.0 & 94.1 &  36.1 &    45.1 &    97.5 &  96.8 & 97.4 &      96.2 & 42.9 \\
   & CmBERT+ptrn IOB2 & 90.4 & 94.6 &  48.3 &    44.8 &    97.5 &  \textbf{98.4} & 97.4 &      98.1 & \textbf{64.1} \\
\bottomrule
\end{tabular}}
\end{table}
\begin{table}[tb]
\centering
\caption{F1 score measured for each approach, pre-trained model and tag format (mean of 5 runs) on the noisy dataset for each entity type.}
\label{tab:classesmetricsocr}
\resizebox{\textwidth}{!}{\begin{tabular}{llrrrrrrrrr}
\toprule
 & Model \& tags &  PER &  ACT &  DESC &  TITREH &  TITREP &  SPAT &  LOC &  CARD &   FT  \\
\midrule
\multirow{4}{*}{M1} & CmBERT IO & \textbf{90.0} & 92.7 &  49.3 &    20.6 &    94.5 &  94.3 & 94.4 &      69.5 & 48.1 \\
   & CmBERT IOB2 & 88.7 & 92.8 &  42.2 &    27.3 &    94.3 &  94.8 & 94.2 &      79.5 & 33.7 \\
   & CmBERT+ptrn IO & 89.3 & 93.3 &  50.9 &    39.7 &    \textbf{95.1} &  \textbf{96.0} & \textbf{95.1} &      77.0 & 50.3 \\
   & CmBERT+ptrn IOB2 & 89.0 & 92.6 &  48.2 &    37.8 &    94.8 &  \textbf{96.0} & 94.8 &      79.4 & 42.3 \\
\cline{1-11}
\multirow{4}{*}{M2} & CmBERT IO & 89.5 & 93.1 &  47.4 &    39.9 &    94.7 &  95.0 & 94.4 &      74.0 & 41.9 \\
   & CmBERT IOB2 & 89.0 & 92.7 &  49.2 &    45.8 &    94.6 &  95.0 & 94.2 &      80.7 & 45.6 \\
   & CmBERT+ptrn IO & 89.0 & \textbf{93.4} &  48.1 &    \textbf{57.0} &    95.0 &  94.8 & 94.9 &      75.6 & 50.7 \\
   & CmBERT+ptrn IOB2 & 87.8 & 91.9 &  39.2 &    40.8 &    95.0 &  95.5 & 94.8 &      78.5 & \textbf{56.5} \\
\cline{1-11}
\multirow{4}{*}{M3} & CmBERT IO & 89.6 & 93.0 &  \textbf{52.3} &    54.6 &    \textbf{95.1} &  95.0 & 94.5 &      79.9 & 53.9 \\
   & CmBERT IOB2 & 88.2 & 91.9 &  43.9 &    38.3 &    94.5 &  95.1 & 94.1 &      76.8 & 42.9 \\
   & CmBERT+ptrn IO & 88.7 & 92.9 &  46.3 &    55.6 &    95.4 &  95.4 & 95.0 &      75.3 & 56.1 \\
   & CmBERT+ptrn IOB2 & 87.6 & 91.8 &  38.3 &    43.7 &    94.8 &  95.7 & 94.6 &      \textbf{82.7} & 52.2 \\
\bottomrule
\end{tabular}}
\end{table}

F1 score measured for each entity type (see tables \ref{tab:classesmetricsref} and \ref{tab:classesmetricsocr}) unsurprisingly show that models perform less on the less represented classes (as \textit{DESC}, \textit{TITREH} or \textit{FT}). We also observed that if some classes are little affected by noise (as \textit{PER}, \textit{ACT} or \textit{LOC}), this is not the case for the \textit{CARDINAL} entity type. Its F1 scores drop at least from 15 percentage points from the ground-truth dataset to the noisy dataset. This can be explained by the similarity of CARDINAL entities with OCR mistakes, which often are numbers.

\subsubsection{Flat-NER experiments results}

We fine-tune CamemBERT and pre-trained CamemBERT on our flat-NER datasets (derived from our nested-NER datasets) using only IO tags. Results are given in table \ref{tab:flatnerresults}. As for nested NER approaches, we observe that performances with the ground-truth dataset are better than those obtained with the noisy dataset. This result shows the impact of noise on the named entity recognition task. Unsupervised pre-training has no significant effect on the results provided with the model fine-tuned on the ground-truth dataset, but it provides better scores on the noisy dataset. Comparison between true flat-NER results and flat-NER equivalent results provided by nested NER approaches shows that fine-tuning models with structured entities does not provide a significant increase of F1-scores.

\begin{table}[tb]
\centering
\caption{F1-score measured on flat-NER datasets (ground-truth and noisy OCR text) with both models (mean of 5 runs).}
\label{tab:flatnerresults}
\begin{tabular}{llr}
\toprule
Dataset & Test &  F1-Score (in \%)        \\
\midrule
\multirow{2}{*}{Ground-truth} & CmBERT &      96.8 \\
    & CmBERT+ptrn &      96.8 \\
\cline{1-3}
\multirow{2}{*}{OCR} & CmBERT &      94.7 \\
    & CmBERT+ptrn &      94.9 \\
\bottomrule
\end{tabular}
\end{table}

\subsection{Qualitative analysis}

\begin{figure}[tb]
    \centering
    \includegraphics[width=\textwidth]{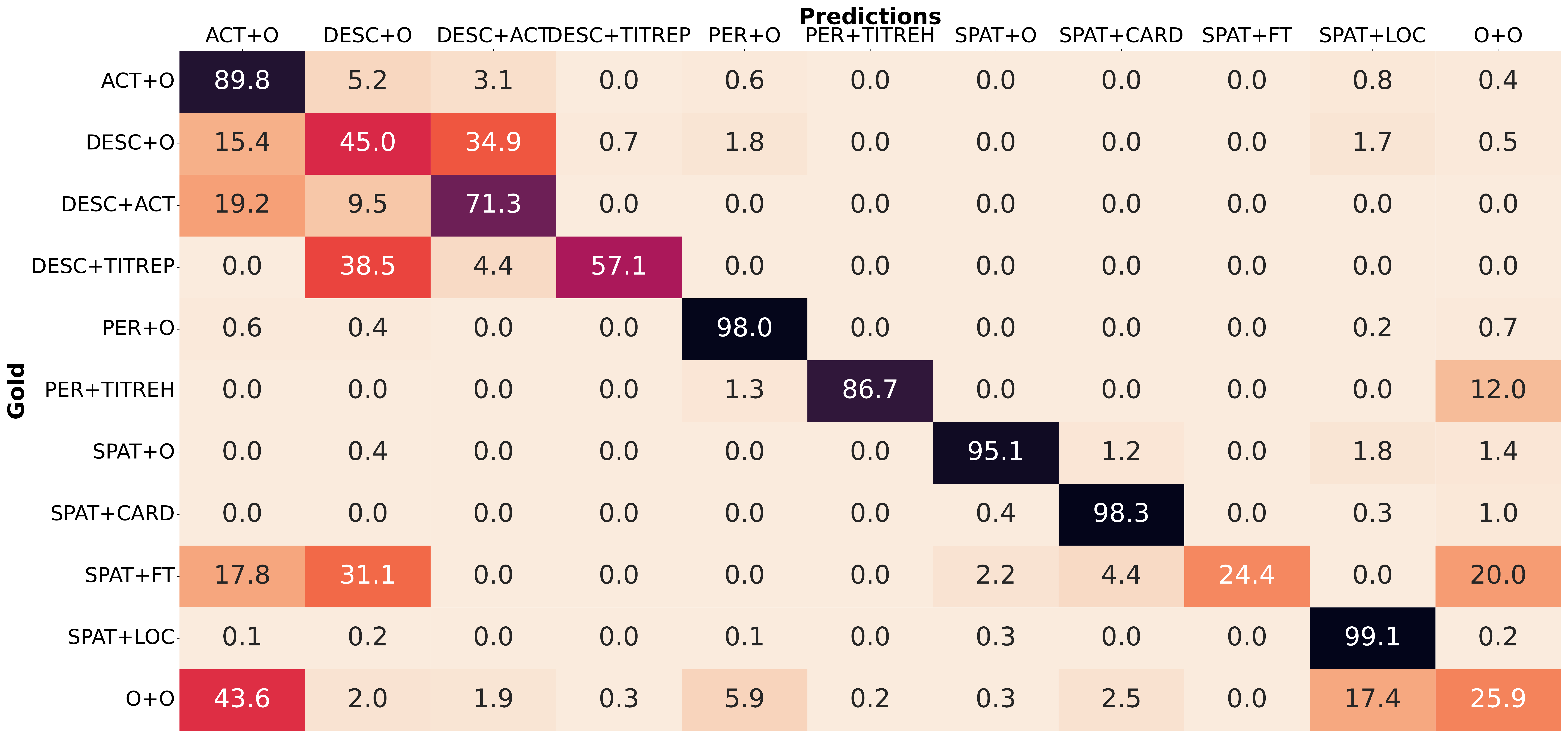}
    \caption{Confusion matrix representing the percentage of tokens associated with each existing joint-label. Values are normalized by row and predictions have been provided by CamemBERT+pre-trained model fine-tuned using our proposed approach [M3] and IO tags.} \label{fig:tokenanalysis}
\end{figure}

In this section, we make a qualitative analysis of our results. We look at the classification errors listing the most common types of errors made by our fine-tuned models. We identify the potential source of errors. Finally, we underline the main pros and cons of each approach.

There are no hierarchy constraints on nested entity types with M1. It produces unauthorised structured entities (as \textit{DESC}/\textit{SPAT} or \textit{SPAT}/ \textit{TITREP} or \textit{ACT/ACT}). This is a significant drawback to producing structured data. The joint labelling based approaches (M2/M3) are a great solution to avoid this issue: hierarchy constraints are defined by the joint labels list set for the training. 

\begin{figure}[tb]
    \centering
    \includegraphics[width=\textwidth]{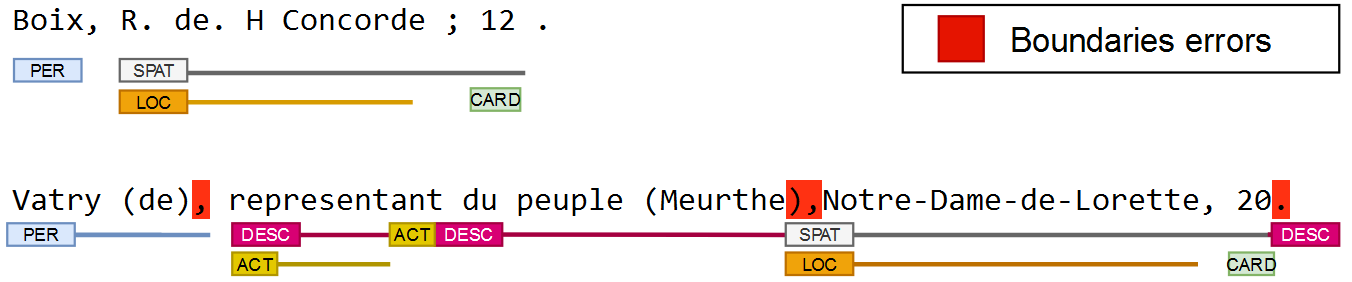}
    \caption{Two noisy entries classified with M3 CamemBERT+ptrn IO model: the top entry doesn't contain mistakes whereas the bottom example contains boundary errors (highlighted) and \textit{DESC/ACT} confusions.}
    \label{fig:ouputs}
\end{figure}

We focus on classification errors made by the M2 and M3 results. They are illustrated by figure \ref{fig:tokenanalysis}. There is a lot of confusion between the \textit{DESC} and \textit{ACT} entities. They contain similar information. There also are confusions between first-level \textit{ACT} and second-level \textit{ACT} entities which give the same kind of information whatever their entity level, as seen in figure \ref{fig:ouputs}. This first group of confusion appears in the top-left corner of figure \ref{fig:tokenanalysis}. A first element of explanation is that information carried by these classes are really close from a semantic point of view which favours such kind of mistakes. \textit{DESC} entity type also is represented by a few examples in datasets. The model often failed to recognise \textit{TITREP} in \textit{DESC} and \textit{TITREH} in \textit{PER}. There is a small number of these entities in the training dataset, which could explain these errors. Note that both titles entity types always contain Unicode characters which aren't included in the CamemBERT model vocabulary.
Confusion also can affect unrelated types of entities. The most often observed errors are \textit{ACT} entities that mention a warehouse of merchandise, which are confused with a \textit{FT} mentioned in an address. The model often misses \textit{FT} type entities in text. For both datasets, we observed that punctuation between entities at the two levels are often misclassified (see the bottom line of figure \ref{fig:tokenanalysis} and highlighted tokens in figure \ref{fig:ouputs}). They are included in their right-side or left-side entity. When using the IOB2 tagging model, we also observed tokens confusions. Both the training and running time of the first approach are multiplied by the number of levels of entities in the dataset. It's not the case for M2 and M3. The first approach requires a post-treatment step to merge predictions made by each flat-NER layer which is time-consuming. M2 and M3 provide full confidence in the hierarchy of entities, unlike M1.

%Utile ????
%Our experiments also have to deal with misfits between flat NER tools and tags used in our  nested NER approaches. \textit{seqeval} Python library is not appropriate for computing metrics for joint labels used in M2 and M3. We had made a difference between the conceptual joint labels and those we implemented, especially for the IOB2 tagging format. On the contrary, the calculation of \textit{L1+L2} metrics with M1 requires saving labels and predictions made in each run and creating \textit{L1+L2} and \textit{P-L1+P-L2} labels in post-processing.

\section{Conclusion}
%\textbf{TODO rappeler approche générale, son intérêt, les contributions et les éléments principaux qui les étayent. Rappeler démarche reproducible/open research avec lien éventuel vers archive anonymisée code + dataset + résultats.}
In this paper, we compare three Nested Named Entity Recognition approaches on noisy and highly structured texts extracted from historical sources: the Paris trade directors of the 19\textsuperscript{th}  century. The aim of this work is to produce a high-level semantic view of the directories' collection. We make a survey of transfer-learning based approaches and select two intuitive ones: the independent NER layers developed by \cite{jia_nested_2021} [M1] and the joint-labelling approach developed by \cite{agrawal_bert-based_2022} [M2]. We suggest an update of this last approach [M3] replacing the loss function of the initial transformer model with a hierarchical loss developed by \cite{bertinetto_making_2020}. It includes the semantic distance between expected and predicted labels. We compare the performance of each approach using a fine-tuned CamemBERT model and a pre-trained and fine-tuned CamemBERT model. We used our new nested NER datasets with clean and noisy directories entries to fine-tune models using IO and IOB2 tagging formats.
The performance of the three approaches achieves high F1 score values with both datasets. M3 outperforms results on hierarchical entities (\textit{L1+L2}) which are the most adapted to produce structured data. Fine-tuning a pre-trained BERT-based model increases performances on noisy entries. Finally, we conclude that the IO tagging format achieves the best performance in most cases. IOB2 tagging format does not improve our results on noisy outputs. Our code, models and datasets are available on OSF\footnote{\url{https://osf.io/w38xm/?view_only=a0deea52fa074b6a8d7d82d077bcc044}} and HuggingFace\footnote{\url{https://huggingface.co/nlpso}}.

\clearpage{}
% Reference will then be sorted and formatted in the correct style.
%
\bibliographystyle{splncs04}
\bibliography{references.bib}

\begin{thebibliography}{10}
\providecommand{\url}[1]{\texttt{#1}}
\providecommand{\urlprefix}{URL }
\providecommand{\doi}[1]{https://doi.org/#1}

\bibitem{abadie_dataset_2022}
Abadie, N., Bacciochi, S., Carlinet, E., Chazalon, J., Cristofoli, P.,
  Duménieu, B., Perret, J.: A dataset of french trade directories from the
  19th century ({FTD}) (2022). \doi{10.5281/zenodo.6394464}

\bibitem{abadie_das_2022}
Abadie, N., Carlinet, E., Chazalon, J., Duménieu, B.: A {Benchmark} of {Named}
  {Entity} {Recognition} {Approaches} in {Historical} {Documents} {Application}
  to 19th {Century} {French} {Directories}. In: Document Analysis Systems: 15th
  IAPR International Workshop, DAS 2022, Proceedings. La Rochelle, France
  (2022). \doi{10.1007/978-3-031-06555-2\_30}

\bibitem{agrawal_bert-based_2022}
Agrawal, A., Tripathi, S., Vardhan, M., Sihag, V., Choudhary, G., Dragoni, N.:
  {BERT}-{Based} {Transfer}-{Learning} {Approach} for {Nested} {Named}-{Entity}
  {Recognition} {Using} {Joint} {Labeling}. Applied Sciences  \textbf{12}(3),
  ~976 (2022). \doi{10.3390/app12030976}

\bibitem{albers_perks_2022}
Albers, T., Kappner, K.: Perks and {Pitfalls} of {City} {Directories} as a
  {Micro}-{Geographic} {Data} {Source}. Collaborative Research Center
  Transregio 190  \textbf{Discussion Paper No. 315} (2022).
  \doi{10.5282/ubm/epub.90748}

\bibitem{alshammari_impact_2021}
Alshammari, N., Alanazi, S.: The impact of using different annotation schemes
  on named entity recognition. Egyptian Informatics Journal  \textbf{22}(3),
  295--302 (2021). \doi{10.1016/j.eij.2020.10.004}

\bibitem{bell_automated_2020}
Bell, S., Marlow, T., Wombacher, K., Hitt, A., Parikh, N., Zsom, A., Frickel,
  S.: Automated data extraction from historical city directories: {The} rise
  and fall of mid-century gas stations in {Providence}, {RI}. PLOS ONE
  \textbf{15}(8),  e0220219 (2020). \doi{10.1371/journal.pone.0220219}

\bibitem{bertinetto_making_2020}
Bertinetto, L., Mueller, R., Tertikas, K., Samangooei, S., Lord, N.A.: Making
  better mistakes: Leveraging class hierarchies with deep networks. In:
  Proceedings of the IEEE/CVF Conference on Computer Vision and Pattern
  Recognition (CVPR) (2020)

\bibitem{dinarelli_models_2011}
Dinarelli, M., Rosset, S.: Models {Cascade} for {Tree}-{Structured} {Named}
  {Entity} {Detection}. In: Proceedings of 5th International Joint Conference
  on Natural Language Processing. pp. 1269--1278. Asian Federation of Natural
  Language Processing, Chiang Mai, Thailand (2011)

\bibitem{dinarelli_tree-structured_2012}
Dinarelli, M., Rosset, S.: Tree-{Structured} {Named} {Entity} {Recognition} on
  {OCR} {Data}: {Analysis}, {Processing} and {Results}. In: Proceedings of the
  {Eighth} {International} {Conference} on {Language} {Resources} and
  {Evaluation} ({LREC}'12). pp. 1266--1272. European Language Resources
  Association (ELRA), Istanbul, Turkey (2012)

\bibitem{ehrmann_named_2021}
Ehrmann, M., Hamdi, A., Linhares~Pontes, E., Romanello, M., Doucet, A.: Named
  entity recognition and classification in historical documents: A survey. ACM
  Computing Survey  (2021), \url{http://infoscience.epfl.ch/record/297355}

\bibitem{finkel_nested_2009}
Finkel, J.R., Manning, C.D.: Nested {Named} {Entity} {Recognition}. In:
  Proceedings of the 2009 {Conference} on {Empirical} {Methods} in {Natural}
  {Language} {Processing}. pp. 141--150. Association for Computational
  Linguistics, Singapore (2009)

\bibitem{jia_nested_2021}
Jia, L., Liu, S., Wei, F., Kong, B., Wang, G.: Nested {Named} {Entity}
  {Recognition} via an {Independent}-{Layered} {Pretrained} {Model}. IEEE
  Access  \textbf{9},  109693--109703 (2021). \doi{10.1109/ACCESS.2021.3102685}

\bibitem{ju_neural_2018}
Ju, M., Miwa, M., Ananiadou, S.: A {Neural} {Layered} {Model} for {Nested}
  {Named} {Entity} {Recognition}. In: Proceedings of the 2018 {Conference} of
  the {North} {American} {Chapter} of the {Association} for {Computational}
  {Linguistics}: {Human} {Language} {Technologies}, {Volume} 1 ({Long}
  {Papers}). pp. 1446--1459. Association for Computational Linguistics, New
  Orleans, Louisiana (2018). \doi{10.18653/v1/N18-1131}

\bibitem{li_survey_2022}
Li, J., Sun, A., Han, J., Li, C.: A survey on deep learning for named entity
  recognition. {IEEE} Transactions on Knowledge and Data Engineering
  \textbf{34}(1),  50--70 (2022). \doi{10.1109/TKDE.2020.2981314}

\bibitem{marinho_hierarchical_2019}
Marinho, Z., Mendes, A., Miranda, S., Nogueira, D.: Hierarchical {Nested}
  {Named} {Entity} {Recognition}. In: Proceedings of the 2nd {Clinical}
  {Natural} {Language} {Processing} {Workshop}. pp. 28--34. Association for
  Computational Linguistics, Minneapolis, Minnesota, USA (2019).
  \doi{10.18653/v1/W19-1904}

\bibitem{martin_camembert_2020}
Martin, L., Muller, B., Suárez, P.J.O., Dupont, Y., Romary, L., de~la
  Clergerie, {\'E}.V., Seddah, D., Sagot, B.: {CamemBERT}: a {Tasty} {French}
  {Language} {Model}. In: Proceedings of the 58th {Annual} {Meeting} of the
  {Association} for {Computational} {Linguistics}. pp. 7203--7219 (2020).
  \doi{10.18653/v1/2020.acl-main.645}

\bibitem{nadeau_survey_2007}
Nadeau, D., Sekine, S.: A {Survey} of {Named} {Entity} {Recognition} and
  {Classification}. Lingvisticae Investigationes  \textbf{30} (2007).
  \doi{10.1075/li.30.1.03nad}

\bibitem{seqeval}
Nakayama, H.: {seqeval}: A python framework for sequence labeling evaluation
  (2018), \url{https://github.com/chakki-works/seqeval}

\bibitem{neudecker2021}
Neudecker, C., Baierer, K., Gerber, M., Christian, C., Apostolos, A., Stefan,
  P.: A survey of {OCR} evaluation tools and metrics. In: The 6th Int. Workshop
  on Historical Document Imaging and Processing. pp. 13--18 (2021)

\bibitem{ramshaw_text_1995}
Ramshaw, L., Marcus, M.: Text chunking using transformation-based learning. In:
  Third Workshop on Very Large Corpora (1995),
  \url{https://aclanthology.org/W95-0107}

\bibitem{santos2019}
Santos, E.A.: {OCR} evaluation tools for the 21st century. In: Proceedings of
  the Workshop on Computational Methods for Endangered Languages. vol.~1
  (2019). \doi{10.33011/computel.v1i.345}

\bibitem{shen_effective_2003}
Shen, D., Zhang, J., Zhou, G., Su, J., Tan, C.L.: Effective {Adaptation} of
  {Hidden} {Markov} {Model}-based {Named} {Entity} {Recognizer} for
  {Biomedical} {Domain}. In: Proceedings of the {ACL} 2003 {Workshop} on
  {Natural} {Language} {Processing} in {Biomedicine}. pp. 49--56. Association
  for Computational Linguistics, Sapporo, Japan (2003).
  \doi{10.3115/1118958.1118965}

\bibitem{wajsburt_extraction_2021}
Wajsbürt, P.: Extraction and normalization of simple and structured entities
  in medical documents. Ph.D. thesis, Sorbonne Université (2021)

\bibitem{wajsburt_effect_2021}
Wajsbürt, P., Taillé, Y., Tannier, X.: Effect of depth order on iterative
  nested named entity recognition models. In: Tucker, A., Henriques~Abreu, P.,
  Cardoso, J., Pereira~Rodrigues, P., Riaño, D. (eds.) Artificial Intelligence
  in Medicine. pp. 428--432. Lecture Notes in Computer Science, Springer
  International Publishing (2021). \doi{\url{10.1007/978-3-030-77211-6_50}}

\bibitem{wang_nested_2022}
Wang, Y., Tong, H., Zhu, Z., Li, Y.: Nested {Named} {Entity} {Recognition}: {A}
  {Survey}. ACM Transactions on Knowledge Discovery from Data  \textbf{16}(6),
  108:1--108:29 (2022). \doi{10.1145/3522593}

\bibitem{zheng_boundary-aware_2019}
Zheng, C., Cai, Y., Xu, J., Leung, H.f., Xu, G.: A {Boundary}-aware {Neural}
  {Model} for {Nested} {Named} {Entity} {Recognition}. In: Proceedings of the
  2019 {Conference} on {Empirical} {Methods} in {Natural} {Language}
  {Processing} and the 9th {International} {Joint} {Conference} on {Natural}
  {Language} {Processing} ({EMNLP}-{IJCNLP}). pp. 357--366. Association for
  Computational Linguistics, Hong Kong, China (2019).
  \doi{10.18653/v1/D19-1034}

\end{thebibliography}
\end{document}